\RequirePackage{amsmath}
\documentclass[copyright,creativecommons]{eptcs}
\usepackage{amsmath}
\usepackage{amssymb,mathrsfs}
\usepackage{graphicx}
\usepackage{epstopdf}	
\usepackage{multirow}
\usepackage[vlined,linesnumbered]{algorithm2e}
\SetAlgorithmName{Strategy}{strategy}{List of Strategies}
\usepackage{rotating}
\usepackage{url}
\usepackage{booktabs} 
\usepackage[caption=false]{subfig}
\newcommand{\X}{\mathcal{X}} 

\newcommand{\ra}{\rightarrow}
\newcommand{\Interface}{\mathit{Interface}}

\newcommand{\PORGY}{{\sc Porgy}}

\newtheorem{definition}{Definition}

\title {Implementation of a Port-graph Model for Finance}
\author  {Nneka Ene\institute{King's College London\\ Dept. of Informatics}\email{nneka.ene@kcl.ac.uk}} 

\def\authorrunning{N. Ene}
\def\titlerunning{Implementation of a Port-graph Model for Finance}

\begin{document}
\maketitle
\begin{abstract}
~In this paper we examine the process involved in the design and implementation of a port-graph model to be used for the analysis of an agent-based \emph{rational negligence} model. Rational negligence describes the phenomenon that occurred during the financial crisis of 2008 whereby investors chose to trade asset-backed securities without performing independent evaluations of the underlying assets. This has contributed to motivating the search for more effective and transparent tools in the modelling of the capital markets.

~This paper shall contain the details of a proposal for the use of a visual declarative language, based on strategic port-graph rewriting, as a visual modelling tool to analyse an asset-backed securitisation market.

\end{abstract}

\section{Introduction}
In this paper we examine the process involved in the design and implementation of a port-graph model to be used for the analysis of an agent-based \emph{rational negligence} model. Rational negligence describes the phenomenon that occurred during the financial crisis of 2008 whereby investors chose to trade asset-backed securities without performing independent evaluations of the underlying assets. This proposal is motivated by the interest an analyst or policy-maker might have in analysing whether or not the purchase of a particular class of asset-backed security ought, going forward, to be subjected to a full due-diligence. 

By replacing more traditional Dynamic Stochastic General Equilibrium (DSGE) models~\cite{dsge} with heterogenous, proactive, agent-based models able to produce more realistic representations, a more efficient system is produced. Such a system is able to support rapid prototyping, run system simulations, and, thanks to its formal semantics, also reason about system properties. Turning to a declarative port-graph transformation system facilitates the analysis of processes of interest. This is because it is able to model the dynamic behaviour of complex systems given that its declarative nature and visual aspects produces a shorter distance between mental picture and implementation. Such a tool is not only able to convert a black box model into a white box but also provide an extensively flexible platform as shall be seen later. There are non-visual elements, such as an inbuilt strategy language, but details of the states produced by the resulting strategy programs are highlighted in a visual trace/derivation tree.

In addition, over extant approaches, the rewrite rules that drive graph transformation systems are an intuitive and natural way of expressing dynamic, structural changes which are generally more difficult to model using traditional simulation approaches where the structure of the model is usually fixed~\cite{DBLP:journals/sosym/LaraGBHT14}. Over many of the declarative languages used within agent-based simulation tools, this approach, given its visual declarative nature, provides more conceptualization support. The use of the port-graph highlights the heterogeneous agent-based nature of the model. Given support for attributes at component-level more realistic simulations can be produced. As a graph transformation system underlying, port-graphs are in addition particularly useful in the development of more concise graph models and in the production of more concise model results given support of both topology and data at the same time. The port-graph makes comprehensive modeling much easier to manage and provides extensive conceptualization support. 

\paragraph{Contributions.}
We outline the typical specification of a graph program that can be used to represent the workings of a small Rational Negligence agent-based system. 
We include broad implementation details highlighting design choices and alternatives that could otherwise have guided our approach. We also carry out a thorough validation of the model with respect to its equational specification producing a base case that provides an effective platform for incrementally increasing the complexity and scope of the model. 
In a previous work; \cite{ahp}, we have defined a hierarchical port-graph extension useful in producing a more integrated model.

\paragraph{Overview.} This paper is organised as follows: We briefly examine the general structure of a simple Port-graph Transformation System and equational semantics of the chosen Rational Negligence model in Section~\ref{section_spec}. Details of the actual implementation can be found in Section~\ref{section_gtsmodel} and Section~\ref{section_testing} examines the results of key tests and checks on the system. We finally conclude and briefly outline future plans in Section \ref{section_conclusion}.

\section{Background}
\label{section_spec}

\subsection{Port-graph Transformation Systems} A port graph is a graph where nodes have explicit connection points, called ports, and edges are attached to ports. Nodes, ports and edges are labelled by a set of attributes describing properties such as colour, shape, etc. Port graphs are transformed by applying port graph rewrite rules. We refer to~\cite{grs_porgy_main_long} for a formal definition of labelled port graphs, where labels are records, i.e., lists of pairs attribute-value. The  values can be concrete (numbers, Booleans, etc.) or abstract (expressions in a term algebra, which may contain variables). More precisely: 

\begin{definition}[Signature]
\label{def:sig}
A port graph signature $\nabla$ consists of   the following pairwise disjoint sets:
$\nabla_\mathscr{A}$, a set of attributes; $\X_\mathscr{A}$, a set of attribute variables;
$\nabla_\mathscr{V}$, a set of values; $\X_\mathscr{V}$, a set of value variables.
\end{definition}

\begin{definition}[Attributed Port Graphs] 
\label{def:portgraphs}
Let $\mathscr{N},\mathscr{P}, \mathscr{E}$ be pairwise disjoint sets. A {\em port graph} over a {\em signature} $\nabla$ is a tuple $G=(V,P,E,D)_{\cal F}$ where $V\subseteq \mathscr{N}$ 
$V$ is a finite set of nodes ($n, n_1, \ldots$ range over nodes); $P\subseteq \mathscr{P}$ 
$P$ is a finite set of ports ($p, p_1, \ldots$ range over ports); $E\subseteq \mathscr{E}$ 
$E$ is a finite set of edges between ports ($e, e_1, \ldots$ range over edges; edges are undirected and two ports may be connected by more than one edge);
$D$ is a set of records over $\nabla$, and ${\cal F}$ is a set of functions $Connect\colon E \ra P \times P$, $Attach\colon P \ra V$ and ${\cal L}\colon V \cup P\cup E \ra D$ such that
\begin{itemize}
\item for each edge $e \in E$, $Connect(e) = (p1,p2)$ the two ports connected by e
\item for each port $p \in P$, $Attach(p) = n$, the node to which p belongs
\item $\mathcal{L}: V \cup P \cup E \rightarrow D$ is the labelling function that returns the record associated to a component
\end{itemize}
Each $n \in V$ maintains an attribute called $\Interface$ whose value is the list of names of the ports attached to $n$, that is, ${\cal L}(n).\Interface = \{{\cal L}(p_i).Name \mid Attach(p_i) = n \}$, 
${\cal L}(n).\Interface = [{\cal L}(p_i).Name \mid Attach(p_i) = n]$, satisfying the following constraint:
${\cal L}(n_1).Name  = {\cal L}(n_2).Name \Rightarrow {\cal L}(n_1).\Interface = 
{\cal L}(n_2).\Interface.$ 
\end{definition}

\begin{definition}[Port-graph Rewrite Rule]
\label{def:pgrr}
A port graph rewrite rule $L \Rightarrow_C R$ is a port graph consisting of two subgraphs $L$ and $R$ together with a node (called arrow node) that captures the correspondence between the ports of $L$ and the ports of $R$, and includes a condition $C$ that will be checked at matching time. More precisely, each of the ports in the arrow node has an attribute Type, which can have three different values: bridge, wire and blackhole, values that indicate how a rewriting step using this rule should affect the edges that connect the redex to the rest of the graph and that satisfy the following conditions:
\begin{enumerate}
\item  A port of type bridge must have edges connecting it to $L$ and to $R$ (one edge to $L$
and one or more to $R$).
\item  A port of type blackhole must have edges connecting it only to $L$ (at least one edge).
\item  A port of type wire must have exactly two edges connecting to $L $ and no edge connecting to $R$.  
\end{enumerate}
\end{definition}

Let $X$ and $Y$ be two port graphs over the same signature $\nabla$. A port graph morphism
$g\colon X \rightarrow Y$ maps nodes, ports and edges of $X$ to those of $Y$ such that the attachment of ports and the edge connections are preserved, and all attributes are preserved except for variables in $X$, which must be instantiated in $Y$ (attribute instantiation can take place within the \emph{algorithm tab} associated with the rule). Intuitively, the morphism identifies a subgraph of $Y$ that is equal to $X$ except at positions where $X$ has variables (at those positions $Y $ could have any value). 

\begin{definition}[Port Graph Morphism] 
More formally given two port graphs $X = (V_X, P_X, E_X, D_X)_{\mathcal{F}_X}$ and $Y = (V_Y, P_Y, E_Y, D_Y)_{\mathcal{F}_Y}$ over the same signature $\nabla$, a morphism f from X to Y, denoted $f:X \rightarrow Y$, and with a definition domain $Dom(f)$, is a family injective functions $\langle f_V:V_X \rightarrow V_Y, f_p:P_X \rightarrow P_Y, f_E:E_X \rightarrow E_Y, f_D:D_X \rightarrow D_Y\rangle$ such that:
\begin{enumerate}
\item $f_V, f_P, f_E$ are injective i.e. distinct components are not identified
\item $\forall e \in E_X:$ if $Connect{_X}(e)=(p_1, p_2)$ then $(f_P(p_1),f_P(p_2))=$\\$Connect{_Y}(f_E(e))$
\item $\forall n \in V_X:$ if $Attach{_X}(p)=n$ for some $p$ then $f_V(n)=Attach_{Y}(f_P(p))$ 
\item $\forall n \in Dom(f), f_D(\mathcal{L}{_X}(n))=\mathcal{L}{_Y}(f_V((n))$\\
$\forall p \in Dom(f), f_D(\mathcal{L}{_X}(p))=\mathcal{L}{_Y}(f_P((p))$\\
$\forall e \in Dom(f), f_D(\mathcal{L}{_X}(e))=\mathcal{L}{_Y}(f_E((e))$
\end{enumerate}
Note that $f_D$ can also instantiate variables
\end{definition}

We denote by $g(X)$ the subgraph of $Y$ consisting of the set of nodes, ports and edges
that are images of nodes, ports and edges in $X$. 

Let $G$ be a port graph. A \emph{rewrite step}  $G \Rightarrow H$ via the port graph rewrite rule $L \Rightarrow_C R$ is obtained by replacing in $G$ a subgraph $g(L)$ by $g(R)$, where $g$ is a morphism from $L$ to $G$ satisfying $C$. More precisely:  

\begin{definition}[Match]\cite{grs_porgy_main_long}
\label{def:match}	
Let $L \Rightarrow R$ be a port graph rewrite rule and $G$ a port graph. We say a match $g(L)$ of the left-hand side (i.e., a redex) is found if: there is a port graph morphism $g$ from $L$ to $G$ (hence $g(L)$ is a subgraph of $G$), $C$ holds, and for each port in $L$ that is not connected to the arrow node, 
its corresponding port in $g(L)$ must not be an extremity in the set of edges of $G - g(L)$.  This last point ensures that ports in $L$ that are not connected to the arrow node are mapped to ports in $g(L)$ that have no edges connecting them with ports outside the redex, thus ensuring that there will be no dangling edges when $g(L)$ is replaced by $g(R)$.
\end{definition}

For a given graph, several outcomes on application of a rule may be possible (due to the intrinsic non-determinism of rewriting). Strategies in  rewriting systems are a means of controlling the creation of rewriting steps and improving rewriting opportunities. A sequence of rewriting steps is called a \emph{derivation}. A \emph{derivation tree} is a collection of derivations with a common root. Intuitively, the derivation tree is a representation of the possible evolutions of the system starting from a given initial state (each derivation provides a trace, which can be used to analyse and reason about the behaviour of system). In \PORGY~\cite{grs_porgy_main_long}, the strategy language allows us to control the way derivations are generated.  
The strategy expression \texttt{setPos(crtGraph)} sets the position graph as the full current graph. If $T$ is a rule, then the strategy $one(T)$ randomly selects one possible occurence of a match of rule $T$ in the current graph $G$, which should superpose the position subgraph $P$ but not superpose the banned subgraph $Q$. This strategy fails if the rule cannot be applied. $Id$ and $Fail$ denote success and failure, respectively. The strategy expression $match(T)$ is used to check if the rule $T$ can be applied (i.e., if there is a match for the left hand side of the rule in the current graph) but does not apply the rule. $(S)orelse(S')$ tries strategy $S$ and if  it fails  then tries to apply $S'$. If both strategies fail then the whole statement fails. $ppick(T_1,\ldots,T_n,\Pi)$ selects one of the transformations $T_1$, \ldots $T_n$ according to the given probability distribution $\Pi$. $while(S)[(n)]do(S')$ executes strategy $S'$ (not exceeding $n$ iterations if the optional parameter $n$ is specified) while $S$ succeeds. $repeat(S)[max~n]$ repeatedly executes a strategy $S$, not exceeding $n$ times. It can never fail (when $S$ fails, it returns $Id$).  



\subsection{The Rational Negligence Model.} As defined in \cite{FIN_ABM_Essex_CRTransferModel_} ``Securitisation is the process of converting cash flows arising from underlying assets or debts/receivables (typically illiquid such as corporate loans, mortgages, car loans and credit cards receivables) due to the originator into a smoothed liquid marketable repayment stream" and this ensures that the originator can raise asset-backed finance through loans or the issuance of debt securities 
also known as assets. An originator is any financial intermediary with a portfolio of assets on its balance sheet. In a securitisation, assets are selected, pooled and transferred to a tax neutral, liquidation-efficient (i.e bankruptcy avoiding), Special Purpose Vehicle (SPV), who funds them by issuing securities. 


In the core rational negligence model~\cite{FIN_agents_absmodel_}, the profit $\mathcal{U}_w$ expected 
by an agent (e.g., a bank) $w$ from trading an asset  depends on whether or not $w$ follows the \emph{negligence rule}, i.e., the rule of not performing independent risk assessment. Let $z$ be a binary variable indicating whether or not the agent is following the negligence rule, then $\mathcal{U}_w(z)$ can be characterised by the following equations, where $p$ is the probability of asset toxicity, $Z$ is the average of all $z$'s in the domain, $c$ is the cost of purchasing an asset (note that the payoff from successfully reselling the asset is normalised to unity), $x_w$ is the cost of performing a complete risk analysis, $k$ is the number of trading partners of the seller bank and $\mathcal{N}_i$ is the set of agents.

\begin{itemize}
\item Expected profit for $w$ when following the negligence rule, i.e., when $z(w) = 1$, if $w$ buys an asset and then tries to sell it to $w'$:
\[\begin{array}{r}
\mathcal{U}_w(1) =^{\textit{def}} -p(1 - z(w'))c + [1 - p(1 - z(w'))](1 - c) \approx 1 - p(1 - Z) - c 
\end{array}\]
This is because if the asset is toxic then $w$ will loose $c$ if $w'$ checks, and will have a profit of $1 -c $ if $w'$ does not check. Of course $w$ does not know a priori whether $w'$ will or not follow the rule, but it can estimate $z(w')$ as the average of all the values of $z$ in the system, $Z$. Note that when $p=0$ the profit is $1-c$ as expected 
\item Similarly, the expected profit for $w$  when the rule is not followed, i.e., $z(w) = 0$, is defined by:
\begin{equation*}
\mathcal{U}_w(0) = (1 - p)(1 - c) - x_w
\end{equation*}
This is because if the asset is toxic, then $w$ will not buy it (losing only $x_w$), but if it is not toxic then it will resell it with a profit of $1-c-x_w$.
\end{itemize}
So the best response of agent $w$ to a buying request is determined by analysing the projected profitability of assuming a negligent approach over a diligent approach:
\begin{equation*}
\mathcal{U}(1)-\mathcal{U}(0) = p(Z-c)+x_w =  p\left( \frac{1}{k} \underset{j \in \mathcal{N}i}{\sum} z_j - c \right) + x_w
\end{equation*}
Following~\cite{FIN_agents_absmodel_}, we implement a model that mimicks the transactions that follow the trading of one asset since this is sufficient to perform validations against equivalent DSGE analyses. The goal of the model is to study the evolution of the system till \emph{fixed point} or \emph{stable state} is reached i.e., in this case, a state such that all potential buyers in the universe of discourse no longer alternate between diligent and non-diligent behaviour in their handling of the purchase of a particular asset. Details of the set of port-graph rules that encapsulate this computation can be found in Tables~\ref{tab:gtsrules}~and~\ref{tab:gtsrules2}. The \emph{beginanalysis} rule in Table~\ref{tab:gtsrules} performs the profitability check.

\section{Implementation}
\label{section_gtsmodel}
We represent the full ABS (Asset-backed Securitization) universe hierarchically as several initial graphs. Port graph rewriting rules and strategies are used to control the step-wise evolution of the graphs and to create a derivation tree that can be used for plotting and analysing parameter values. The asset trading model sits at the top level of the model hierarchy and it is non-deterministic in nature. 
Below this system, also able to handle asset pricing and valuation issues, lie several deterministic subsystems that model origination, structuring of the deal, SPV transfers and profitability of the sale.  
\begin{figure}
	\centering	
	\includegraphics[width=0.8\textwidth]{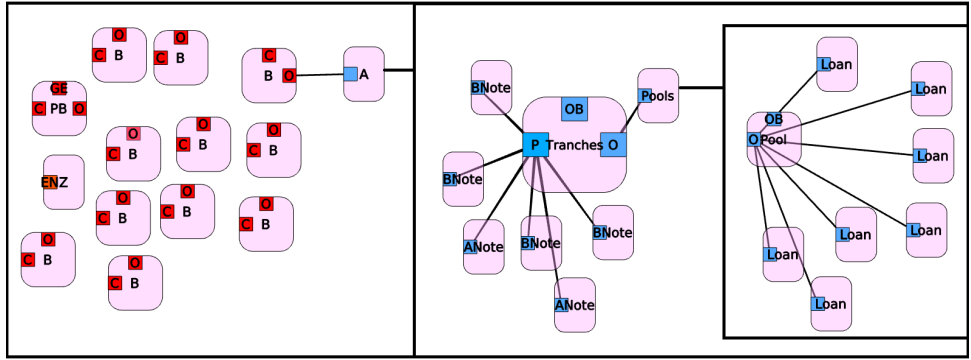}
	\caption{\small \sl All Tiers Flattened and Condensed}\label{figure:all_tiers}
\end{figure}

We chose to model asset-transfer transactions using a combination of global and local data, and a global state, the $Z$ node (an indicator of market behaviour obtained as the average value of each individual bank's approach, represented by the local lower-case attribute $z$ and not to be confused with the global value $Z$). A $\mathit{Change}$ indicator node is used within a rule to detect whether the market has reached a stable state. Tables~\ref{tab:agentnodes} and \ref{tab:nodeports} contain a description of the nodes used. Alternative designs are possible, highlighting the flexibility of the approach: For example, local copies of the z-attribute could be used to propagate negligent/diligent behaviour using propagation algorithms borrowed from  social network models whereby information is transmitted, or in this case received, based on the actions of neighbours or neighbours-of-neighbours or other clusters~\cite{ValletKPM15}. The details of this alternative model shall be future work.

\begin{table}
\centering
\scalebox{1.02}{\begin{tabular}{|l|l|l|}
 \hline
 \multirow{1}{*}{\textbf{Name of Rule}} &~~~~~~~~~~~~~~~~~~~~~~~\textbf{Description}\\  
  \hline
	\multirow{4}{*}{requesttobuy} & Sends a request-to-buy message to a random bank $B$ changing \\
								  & the name of this node to $PB$ (\emph{PotentialBuyer}).\\ 
								  & A copy of the rule:\\
								  & \includegraphics[scale=0.23]{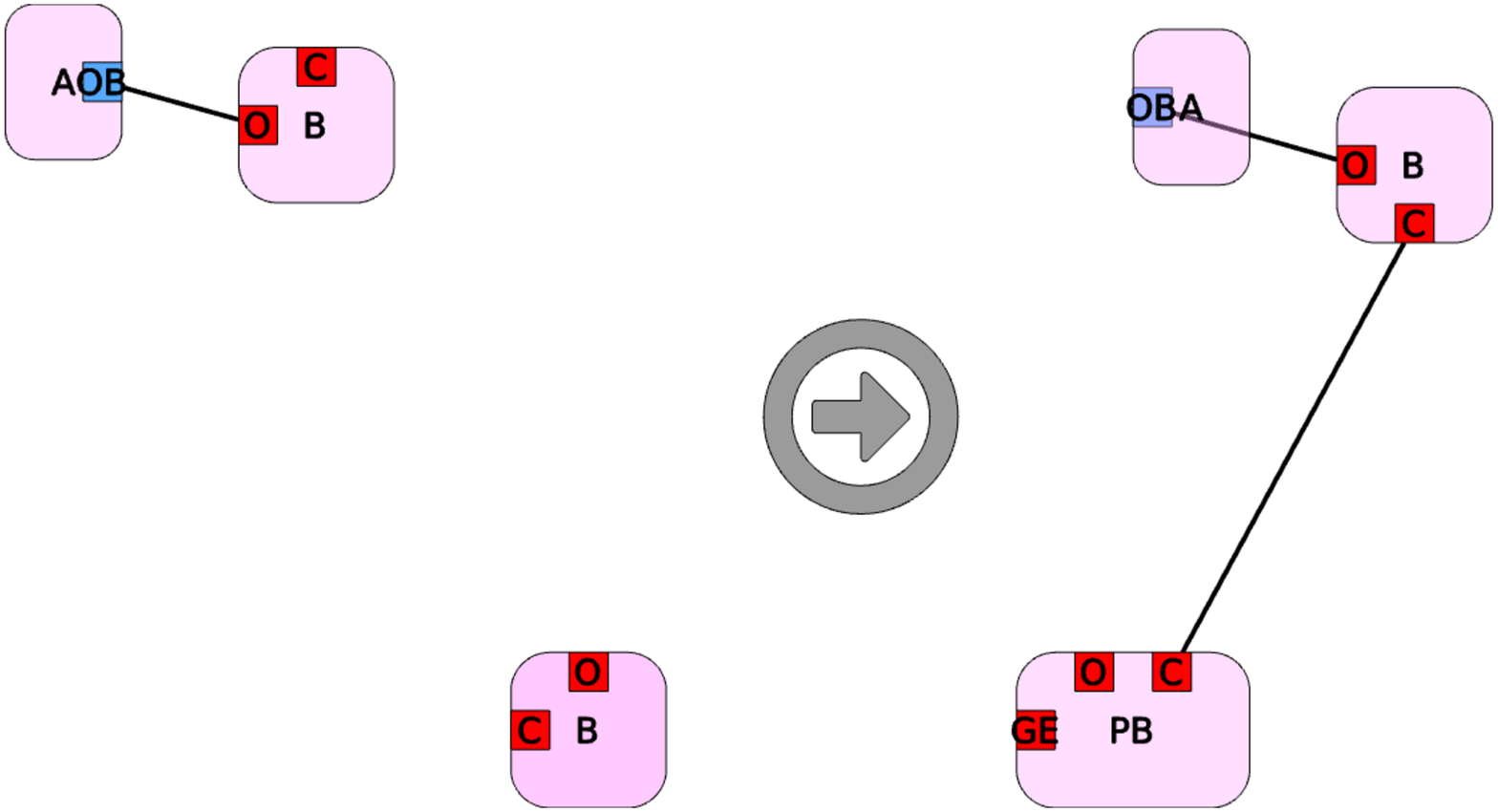}\\
	\hline
	\multirow{7}{*}{beginanalysis}& Computes profitability $\mathcal{U}(1)$, $\mathcal{U}(0)$ of $PB$,  generating \\ 
								  & a node \emph{Theta} with attribute \emph{DeltaU1U0} = $\mathcal{U}(1)-\mathcal{U}(0)$. \\
								  & \includegraphics[scale=0.3]{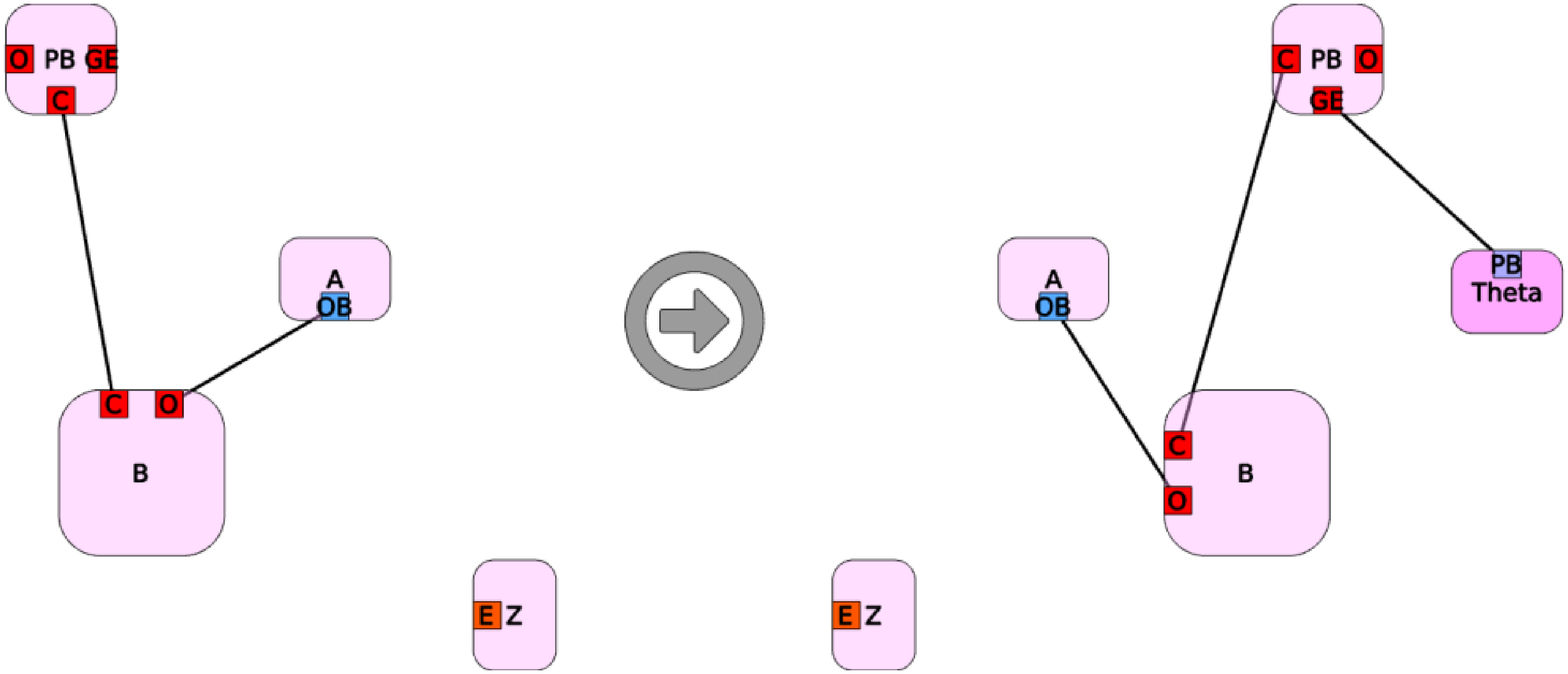}\\
& Computation in the algorithm tab is as follows:\\
& $Theta.U1=1-A.\mathit{p\_tox}(1-Z.z)-A.c\_val$\\
& $Theta.U0=(1-A.\mathit{p\_tox})(1-A.c\_val)-A.\mathit{ddcost}$ \\
& $Theta.\mathit{DeltaU1U0}=Theta.U1-Theta.U0$\\
	\hline
	\multirow{7}{*}{updatez}      & Updates the attribute $Z$ in node Z. The new value in \\	
								  & $Z$ is  $(Z*(k-1)+z(PB))/k$. \\
								  &\includegraphics[scale=0.22]{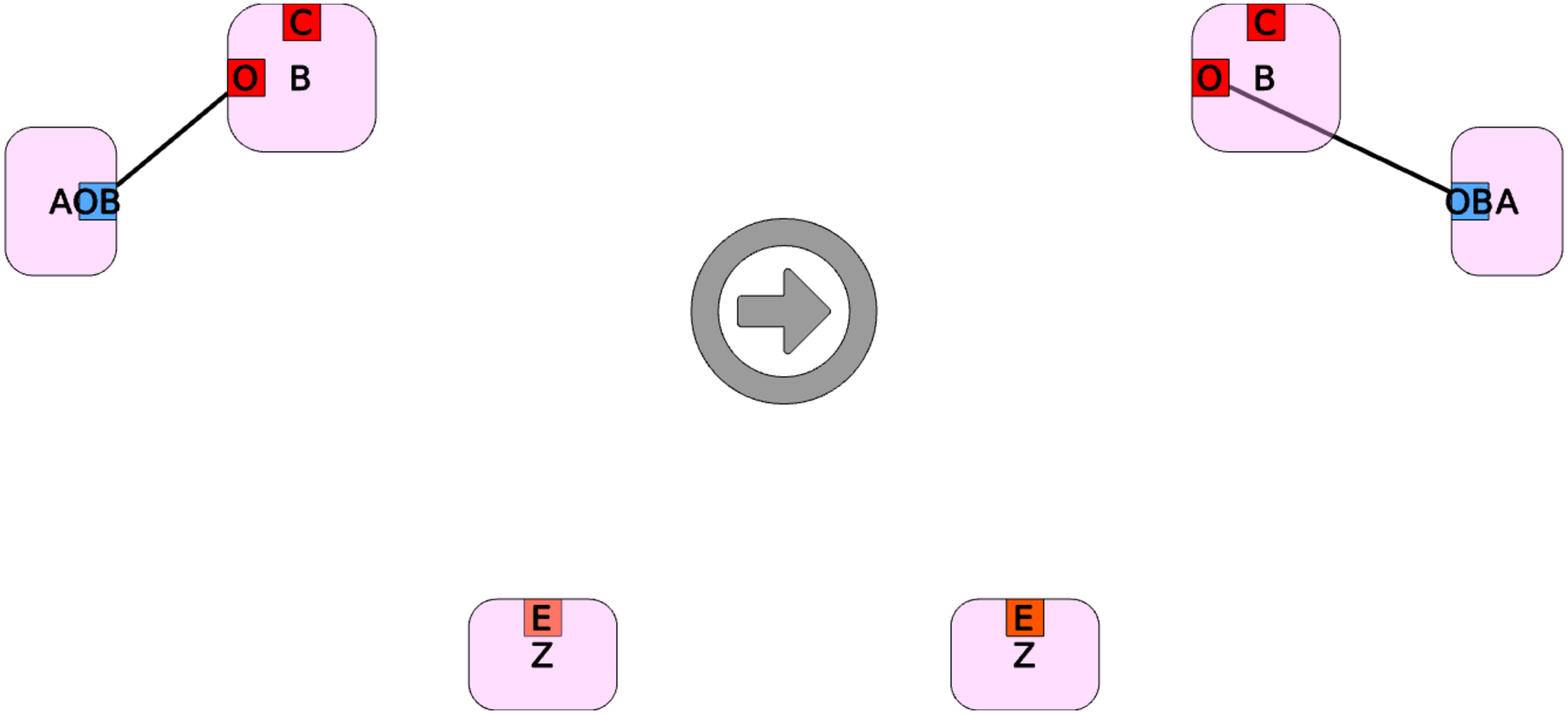}\\
								  & Computation in the algorithm tab is as follows:\\
		    &$Z.z=((Z.z*(Z.\mathit{numofagents}-1))+B.z)/Z.\mathit{numofagents}$\\ 
	\hline	
    \end{tabular}}	
\caption{\small \sl Rewrite Rules}\label{tab:gtsrules}
\end{table}

\begin{table}
\centering
\scalebox{1.02}{\begin{tabular}{|l|l|l|}
 \hline
 \multirow{1}{*}{\textbf{Name of Rule}} &~~~~~~~~~~~~~~~~~~~~~~~\textbf{Description}\\  
  \hline
	\multirow{6}{*}{followresult} & Applies if \emph{DeltaU1U0} $\geq 0$. \\
	                              & As additional visualisation support, it generates a \emph{follow}\\ 	
								  &  node if more profitable to not do a full risk analysis.\\
								  & \includegraphics[scale=0.15]{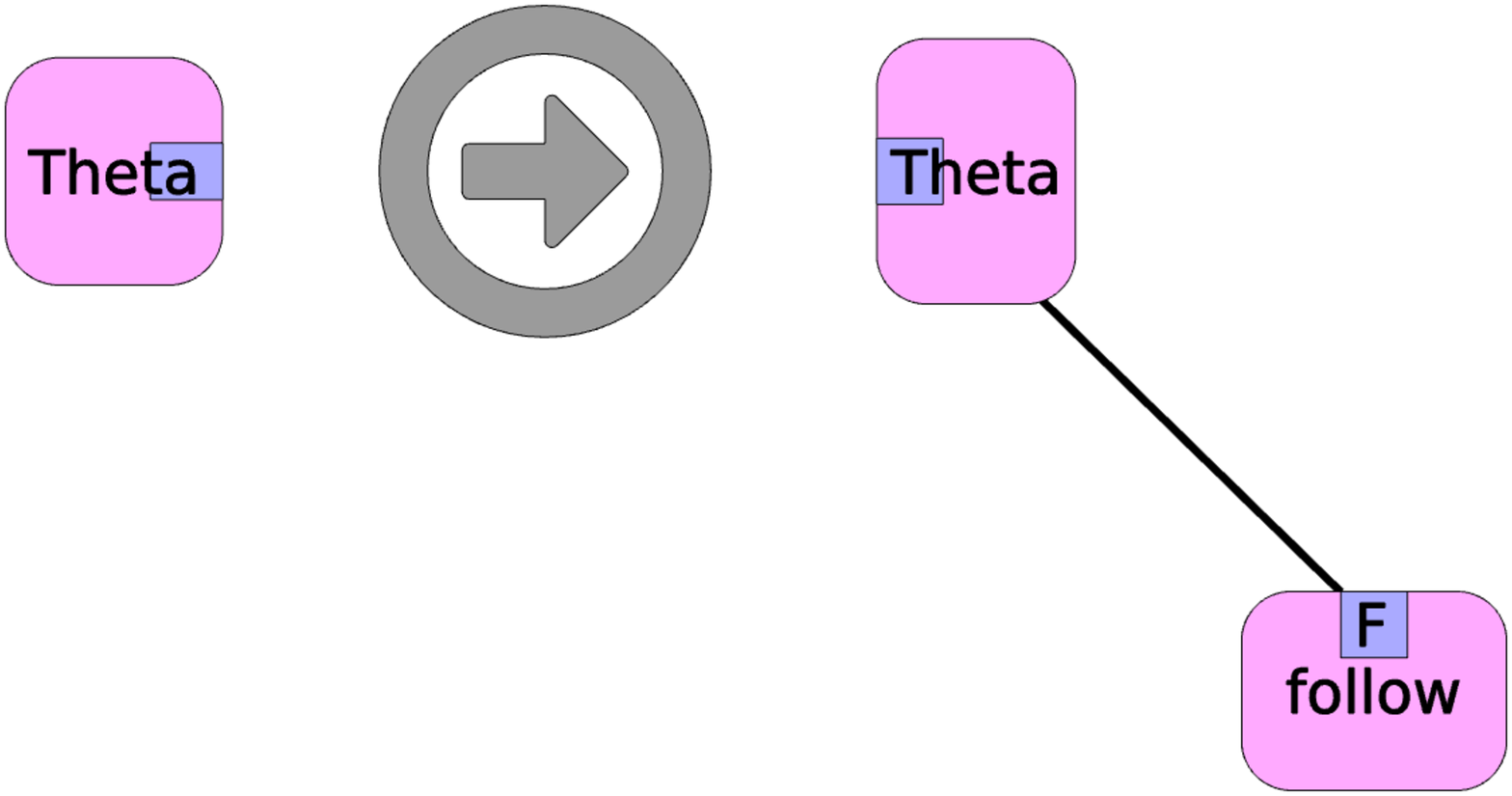}\\
								  & Arrow-node Application Condition:\\
								  & If $\mathit{Theta.DeltaU1U0} \geq 0$\\
	\hline
	\multirow{4}{*}{deviationresult}& Applies if \emph{DeltaU1U0} $< 0$\\
	                                            & As  additional visualisation support, it generates a \emph{deviation} \\ 	
								  & node if more profitable to do a full risk analysis\\
								  & (Similar to \emph{followresult}).\\
	\hline
	\multirow{4}{*}{followdecision} & Transfers asset and prepares for a new transaction (i.e. cleans \\ 	
								  & up after the decision negligence rule), updating bank's attribute $z$, \\
								  & updating the \emph{Change} counter if necessary.\\
								  &\includegraphics[scale=0.3]{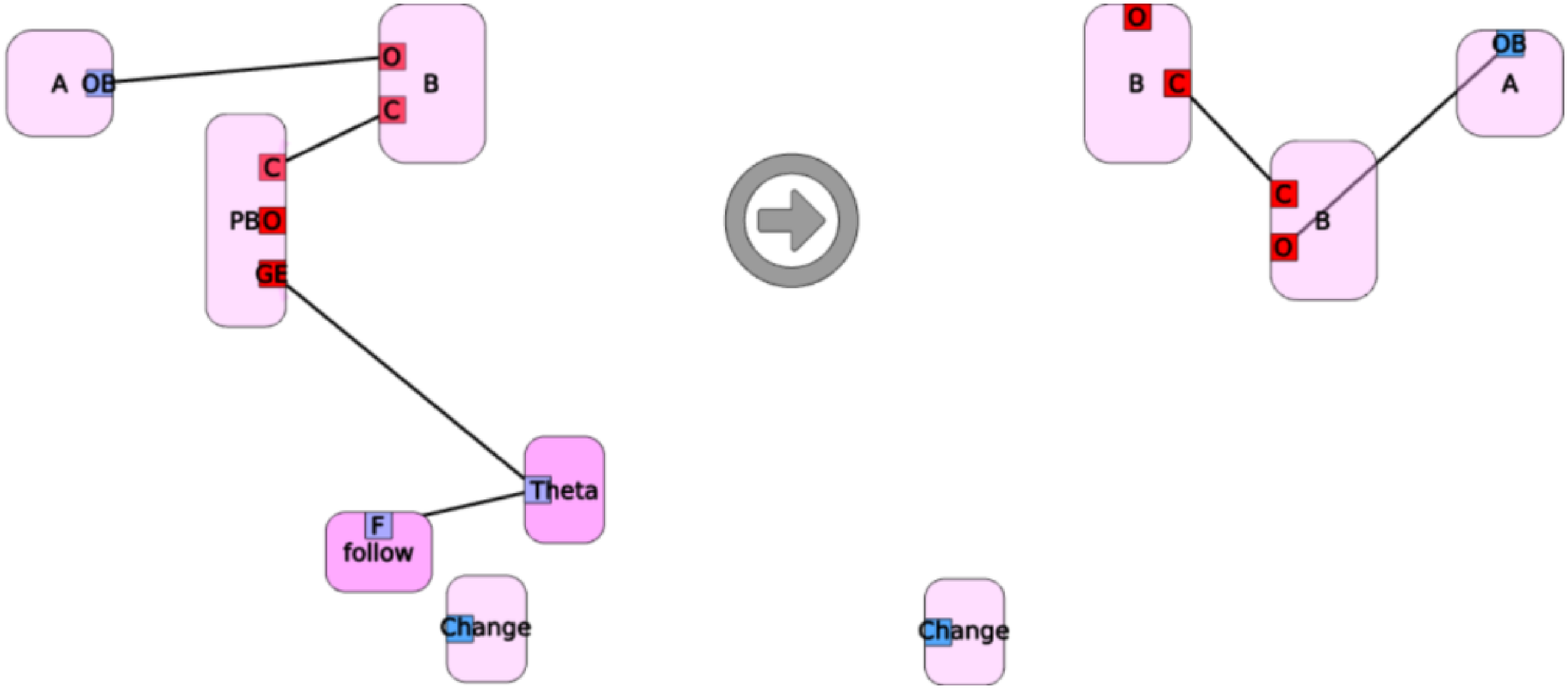}\\
	\hline
	\multirow{4}{*}{deviationdecision} & Transfers asset and prepares for a new transaction (i.e. cleans \\
								  & up after the decision to deviate from the negligence rule), updating \\
								  & bank's attribute $z$, updating the \emph{Change} counter if necessary\\
								  &(Similar to \emph{followdecision}).\\
	\hline
	 \multirow{5}{*}{change} 	  & Sets the \emph{Change} counter back to 0 if greater than 0.\\
	 &\includegraphics[scale=0.10]{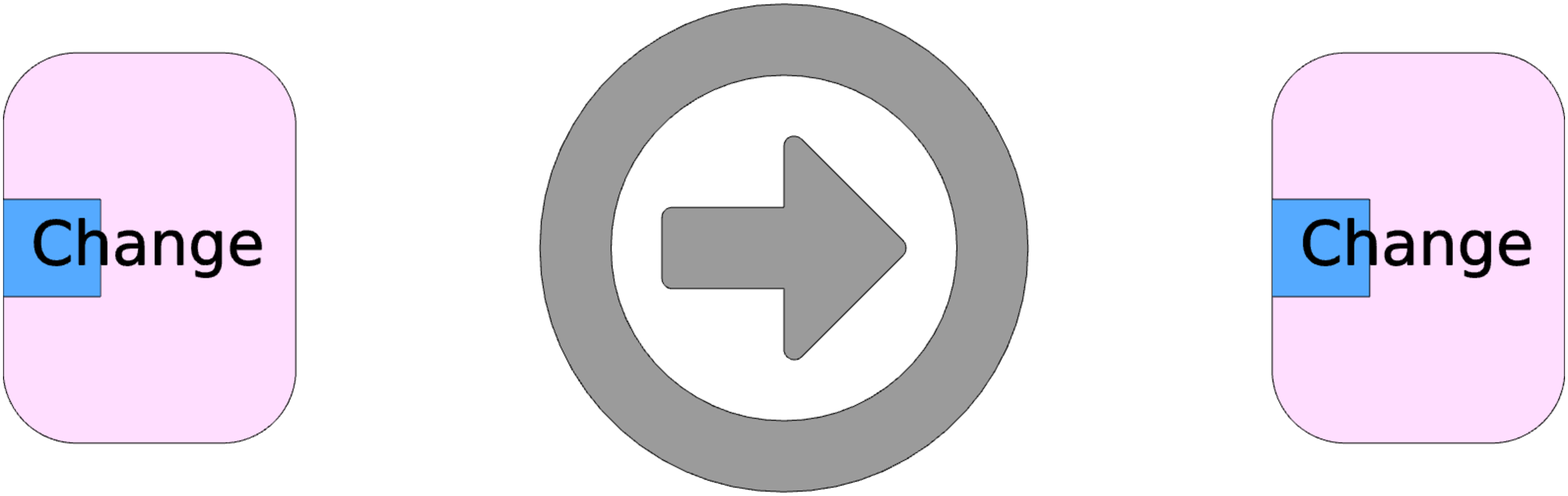}\\
	 &Computation in the algorithm tab is as follows:\\
	 &$\mathit{Change.change}=0$\\
	 &$\mathit{Change.sumofchange}=0$\\
	 \hline
    \end{tabular}}	
\caption{\small \sl Rewrite Rules}\label{tab:gtsrules2}
\end{table}

Tables~\ref{tab:agentnodes} and~\ref{tab:nodeports} describe the nodes in the system and their ports respectively.
\begin{table}
\centering
\scalebox{0.88}{\small{
\begin{tabular}{|l|l|l|}
\hline
\bfseries{Entity} & \bfseries{Attribute} & \bfseries{Description}\\  
\hline

\multirow{5}{*}{Buyer (B/PB)} & Payoff (payoff)  & {Returns from re-selling an asset}.\\
 {Bank/Potential} &  & \\
 \cline{2-3}& $z$ & {Indicates whether or not,}\\
			      &~&{as a rule, the institution}\\			      
				      &~&{performs independent risk analyses}\\
      \cline{2-3}
					     & Bank ID (b\_id)& {Bank identifier}\\
\hline
\multirow{7}{*}{Asset} & Current Value (c\_val) & Cost of purchasing an asset\\
\cline{2-3}
		                     	& Probability   & An asset is toxic if the borrowers \\ 								 
			&  of Toxicity (p\_tox)                         & of the underlying loans are \\          
			    							& 						&  likely to default or are in default\\
       \cline{2-3}
						 & Actualised Toxicity & 		Current toxicity level\\
						 & (a\_tox)& \\

       \cline{2-3}
							 & Perception (pe) & External rating of the asset \\
							&& by rating agencies\\
      \cline{2-3}
					     & Due Diligence Cost &Full cost of an independent  \\
					     						 & (ddcost)& risk assessment\\
\hline
\multirow{2}{*}{Change} & change & Change in bank approach\\
      \cline{2-3}
					     & Sum of change & Sums all changes in a current cycle\\ 
					     & (sumofchange) & \\
\hline
 & 			z   & Represents the global average z\\				
\cline{2-3}
& 			Number of Iterations   & Counter that keeps track of \\	
{Z}& 			(numofiterations)& AllTrade iterations\\	
\cline{2-3}
& 			Number of Agents   & Variable that keeps track of  \\ 
&			(numofagents)  	& number of banks\\	

\hline
\multirow{3}{*}{Theta} & U1 & Profitability of being negligent\\	
\cline{2-3}		
& U0 & Profitability of being diligent.\\				
\cline{2-3}		
& DeltaU1U0 & Difference between U1 and U0.\\				
\hline
\end{tabular}}
}
\caption{\small \sl Nodes and Attributes}\label{tab:agentnodes}
\end{table}

\begin{table}
\centering
\scalebox{0.88}{\small{
\begin{tabular}{|l|l|l|}
\hline
\bfseries{Entity} & \bfseries{Ports} & \bfseries{Description} \\  
\hline
\multirow{2}{*}{Bank} &       O (Owns)                & {Edges attached to this port highlight }\\
&&assets owned by the bank\\
\cline{2-3}
				      & C (Contacts)       & {Communication channel with another bank}\\
\hline
\multirow{1}{*}{Asset} & OB (Owned\_by) & {Connects the asset to its current owner}\\
\hline
\multirow{1}{*}{Z} & EN (Environment) & Global entity that tracks current average sentiment\\				
\hline
\multirow{3}{*}{PotentialBuyer} &  O (Owns)       & {Links to assets owned by the bank}\\
\cline{2-3}
&   C (Contacts)       & {Communication channel with another bank}\\
\cline{2-3}
&   GE (Generates)      & {Declares a relationship with an analysis node}\\
\hline
\multirow{1}{*}{Change} & CH (change) & Counter that keeps track of behaviour changes\\				
\hline
\multirow{1}{*}{Theta} & PB (Produced\_by) & Entity that produces this computation helper\\				
\hline
\end{tabular}}}
\caption{\small \sl Ports in each kind of node}\label{tab:nodeports}
\end{table}

After the creation of a comprehensive set of rules, reduction strategies were created that defined the sub-graphs to be selected for evaluation and which rules should be applied to the starting state of the model and it is from this point that the derivation tree begins to undergo construction as the execution strategy calls on rules that create step-wise transformations. Specifically, the asset transfer processes are governed by the strategies \emph{AllTrade} and \emph{FixedPointSearch} (see Strategies~\ref{alg:fixedpointsearch} and~\ref{alg:alltrade} below), using 8 rewrite rules summarised in Tables~\ref{tab:gtsrules} and~\ref{tab:gtsrules2}.


Also highlighting flexibility is the fact that a variant of strategy \emph{AllTrade} can simply replace the \texttt{orelse} operator by a  \texttt{ppick} operator, and then begin to model probabilistic choice of logit type between following or deviating from the negligence rule. The probability distribution used in this case implements the stochastic ``trembles'' described in~\cite{fin_agents_abslink} and can be written within our strategy environment as follows:
\begin{verbatim}
ppick(followResult, deviationResult, udfLogitModel)
\end{verbatim}
where \texttt{udfLogitModel} is a function that reads the profitability of being negligent or diligent (attributes U1 and U0 in the node Theta of the graph produced by the relevant rule) and returns the following values as a list: 
$$\frac{\exp^{\mathcal{B}U_{i}(z=1)}}{\exp^{\mathcal{B}U_{i}(z=1)} + \exp^{\mathcal{B}U_{i}(z=0)}} ~~~ and ~~~ 1-(\frac{\exp^{\mathcal{B}U_{i}(z=1)}}{\exp^{\mathcal{B}U_{i}(z=1)} + \exp^{\mathcal{B}U_{i}(z=0)}})$$ 
where $i$ is the current agent number and $\mathcal{B}$ is the intensity of choice parameter that controls the ease by which fixed point is reached (as specified in \cite{fin_agents_abslink}). 

\begin{algorithm}[t!]
\SetAlgoLined
\#AllTrade\#;\\
while(match(change))do(\\
~~one(change);\\
~~\#AllTrade\#)\\
   \caption{\small \sl FixedPointSearch}\label{alg:fixedpointsearch}
\end{algorithm}

\begin{algorithm}[t!]
\SetAlgoLined
setPos(crtGraph);\\
repeat(one(requesttobuy); \\
~~one(beginanalysis); \\
~~(one(deviationresult);one(deviationdecision)) orelse\\ 
~~(one(followresult);one(followdecision))\\
~~setPos(crtGraph);\\
~~one(updatez))(k)\\
   \caption{\small \sl AllTrade}\label{alg:alltrade}
\end{algorithm}

\section{Testing}
\label{section_testing}
\begin{figure}[h!]
\centering
\subfloat[Low Toxicity, High Due Diligence Cost,Mixture of Diligent and Negligent Banks]{\includegraphics[width=.25\textwidth]{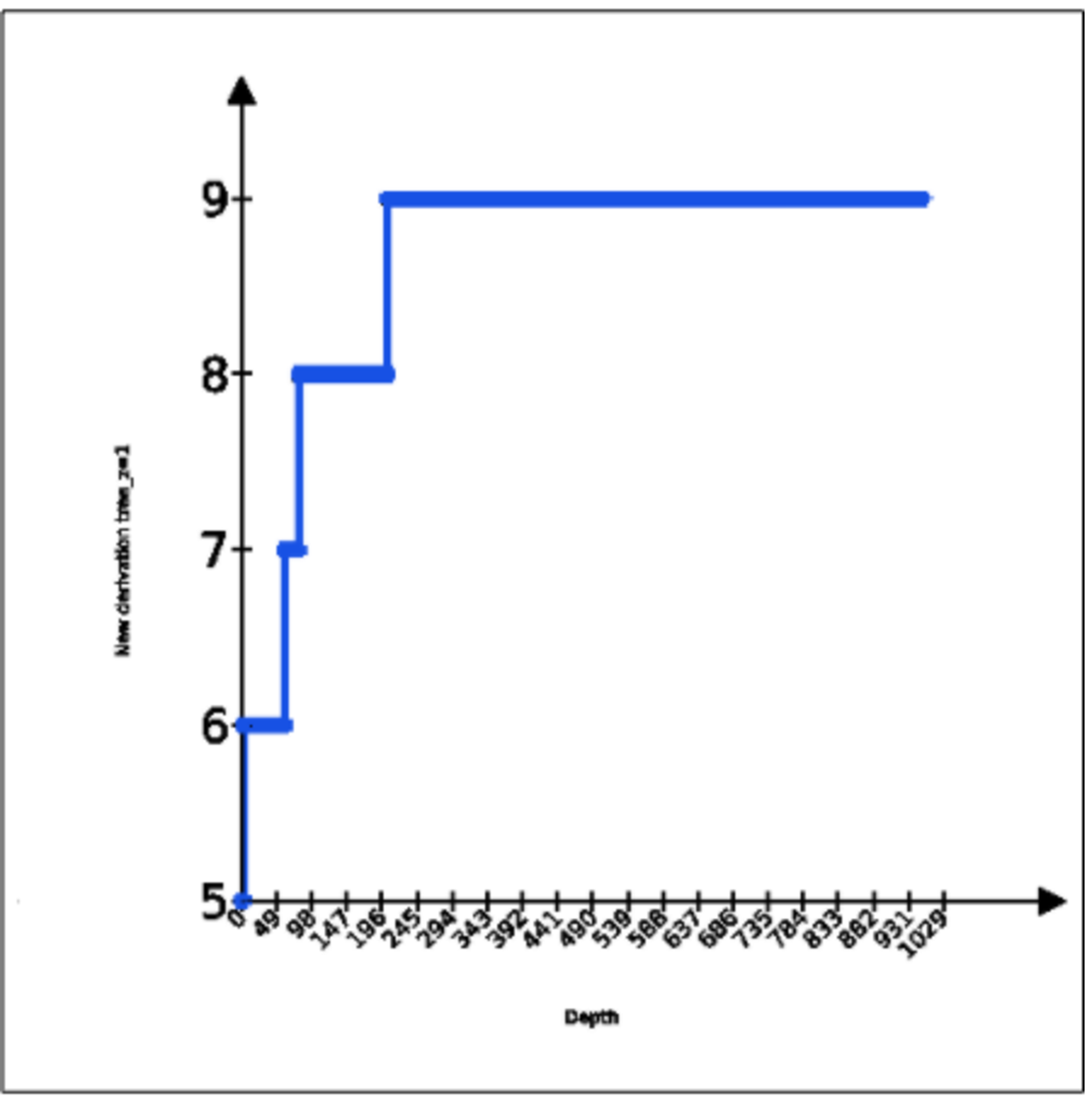}\label{figure:m0545_ptox0001_ddcost01}}
\subfloat[High Toxicity, High Due Diligence Cost, Mixture of Diligent and Negligent Banks]{\includegraphics[width=.25\textwidth]{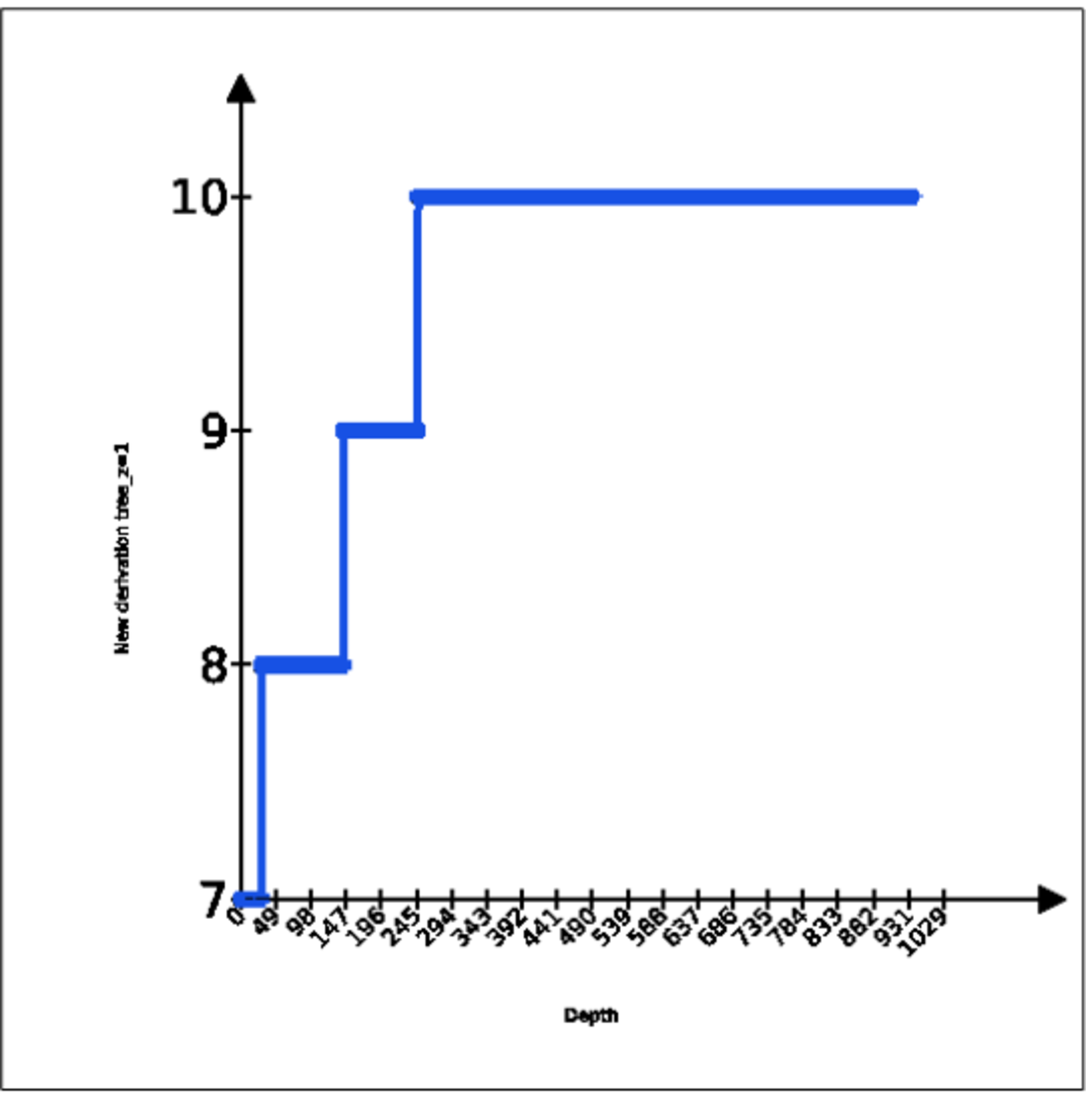}\label{figure:m0545_ptox01_ddcost01}}
\subfloat[High Toxicity, Low Due Diligence Cost, Mixture of Diligent and Negligent Banks]{\includegraphics[width=.25\textwidth]{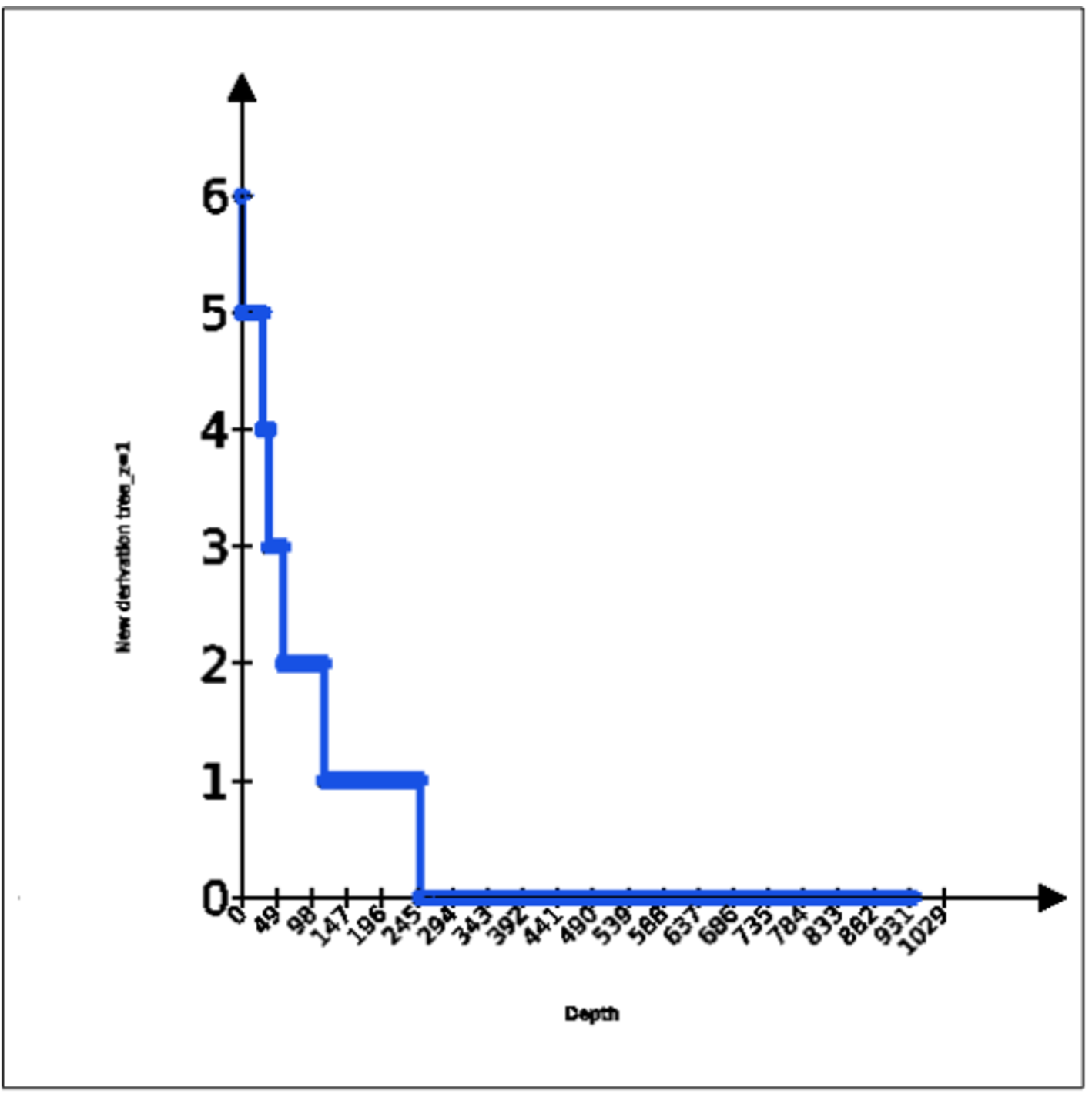}\label{figure:m0545_ptox01_ddcost0001}}
\subfloat[Low Toxicity, Low Due Diligence Cost, Mixture of Diligent and Negligent Banks
]{\includegraphics[width=.25\textwidth]{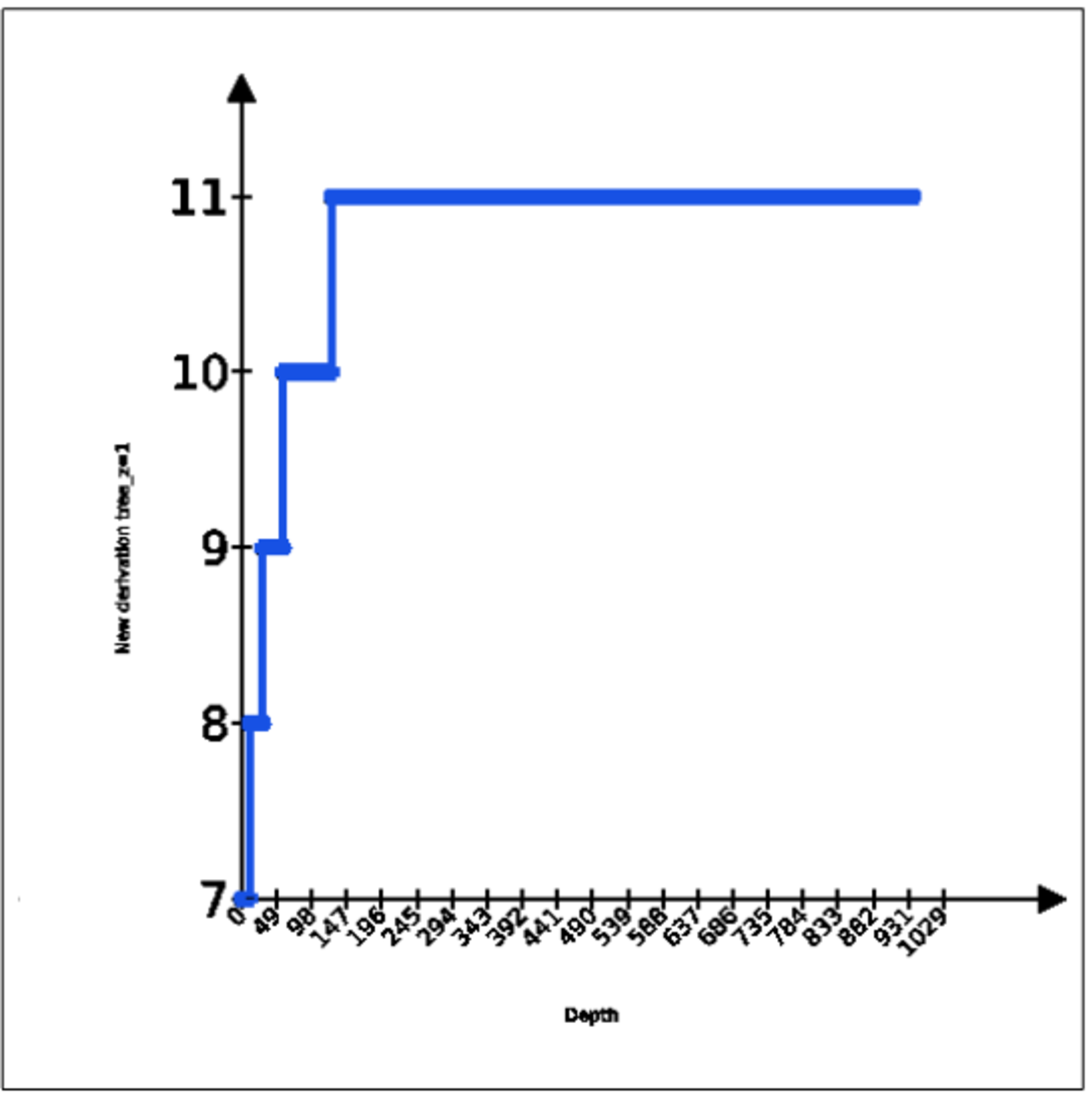}\label{figure:m0545_ptox0001_ddcost0001}}\\
\subfloat[High Toxicity, High Due Diligence Cost, Diligent Banks]{\includegraphics[width=.25\textwidth]{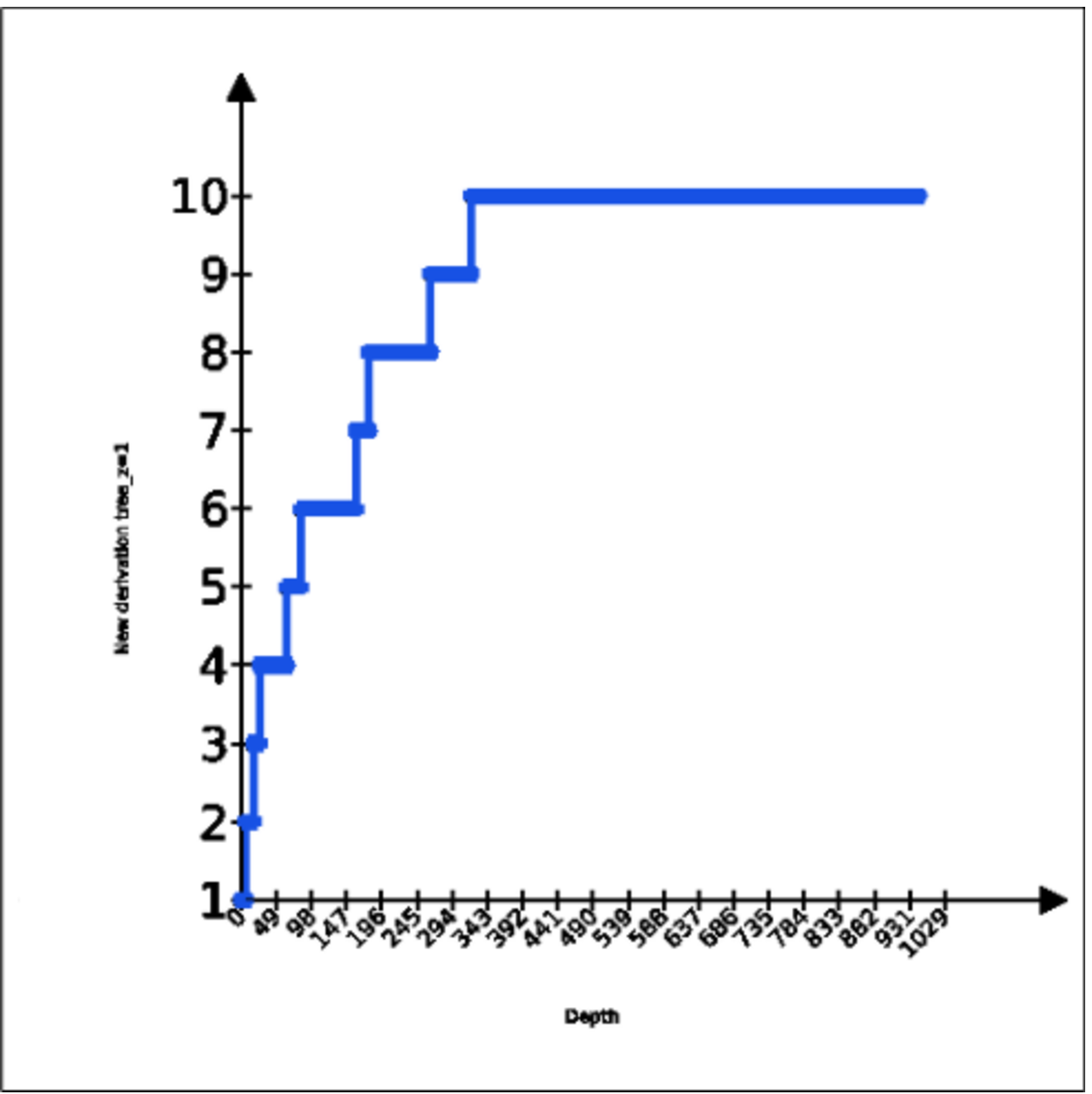}\label{figure:m0_ptox01_ddcost01}}
\subfloat[Low Toxicity, High Due Diligence Cost, Diligent Banks]{\includegraphics[width=.25\textwidth]{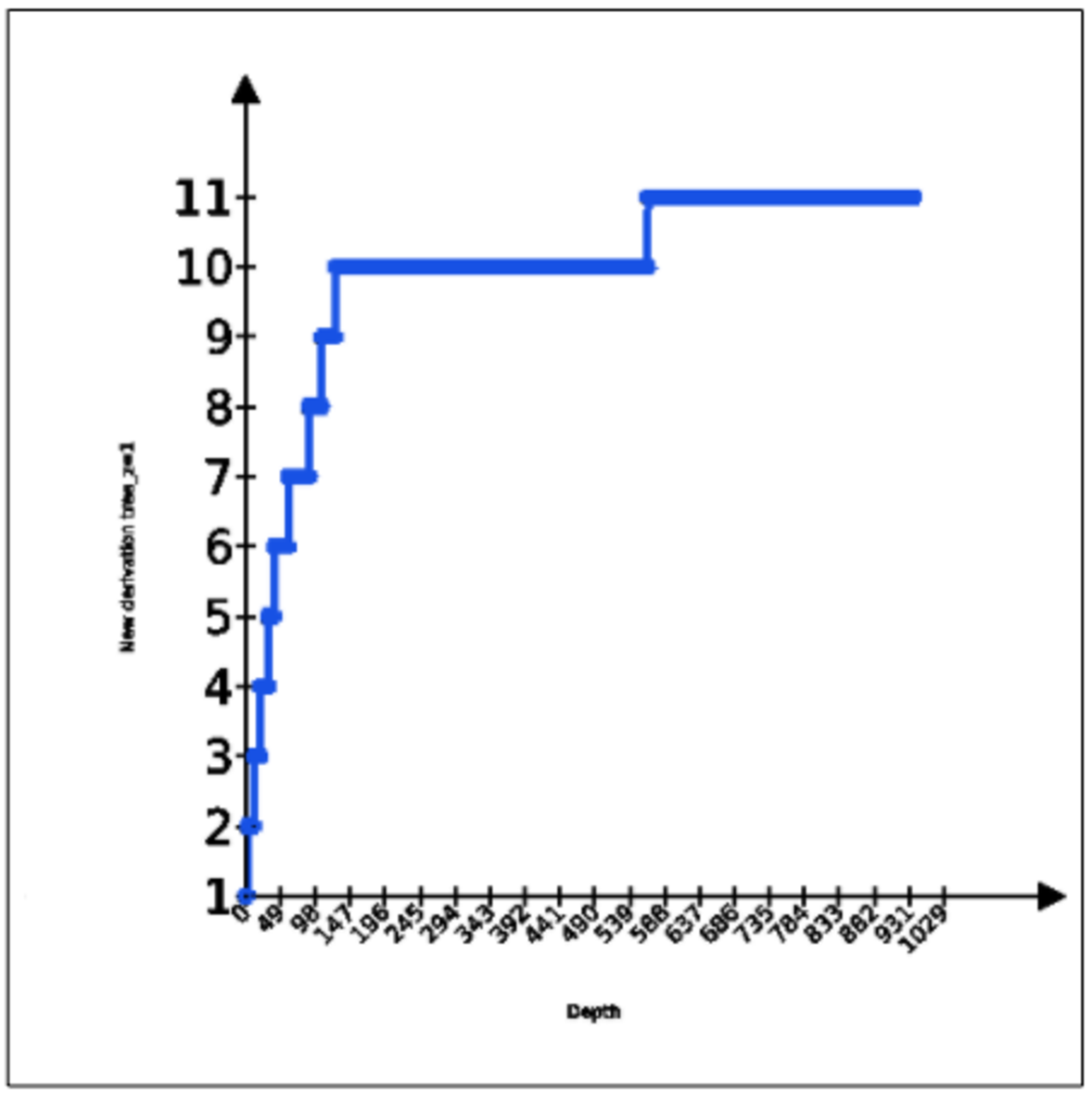}\label{figure:m0_ptox0001_ddcost01}}
\subfloat[Low Toxicity, Low Due Diligence Cost, Diligent Banks]{\includegraphics[width=.25\textwidth]{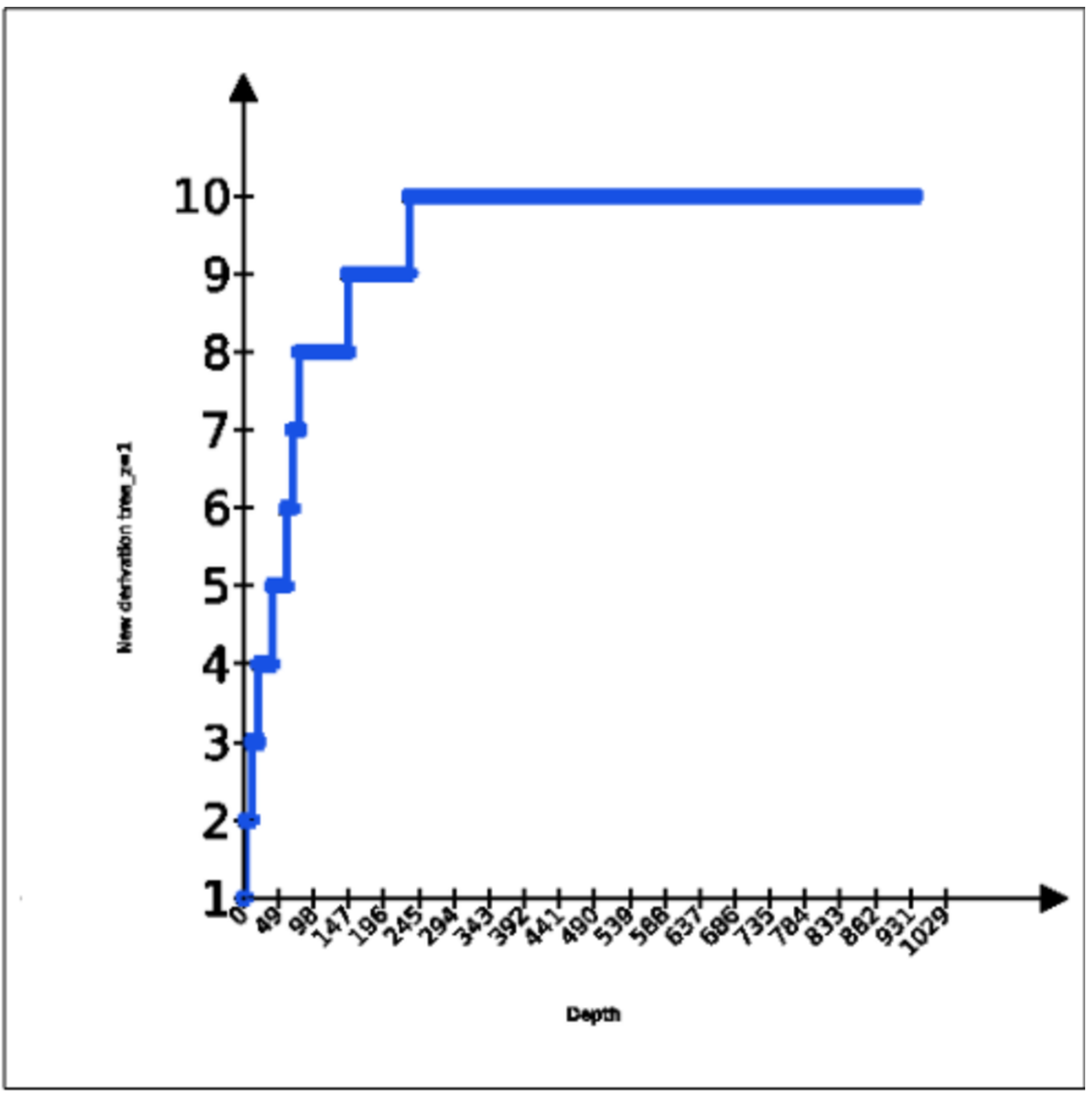}\label{figure:m0_ptox0001_ddcost0001}}
\subfloat[High Toxicity, Low Due Diligence Cost, Diligent Banks]{\includegraphics[width=.25\textwidth]{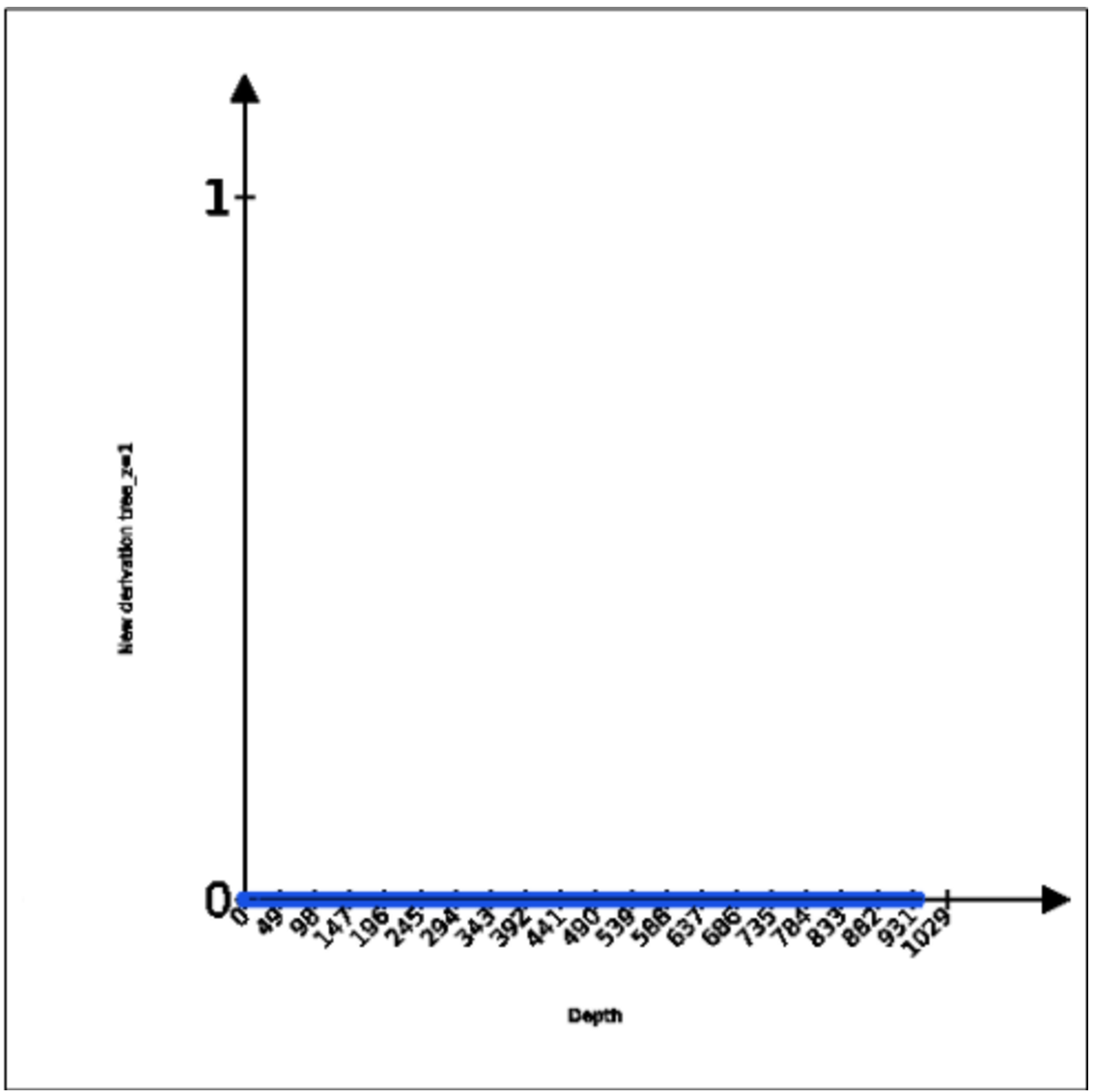}\label{figure:m0_ptox01_ddcost0001}}\\
\subfloat[Low Toxicity, High Due Diligence Cost, Negligent Banks]{\includegraphics[width=.25\textwidth]{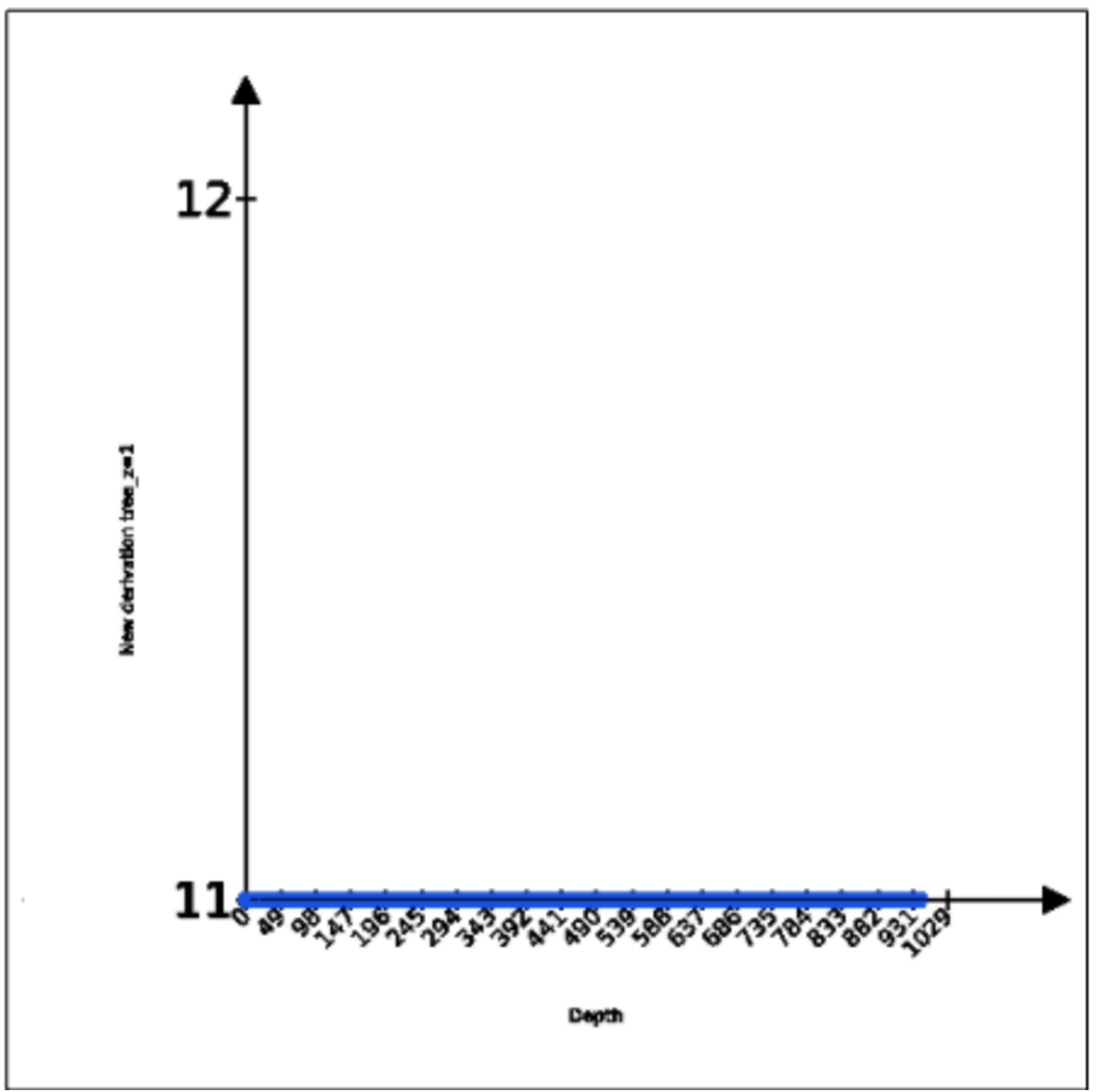}\label{figure:m1_ptox0001_ddcost01}}
\subfloat[High Toxicity, High Due Diligence Cost, Negligent Banks]{\includegraphics[width=.25\textwidth]{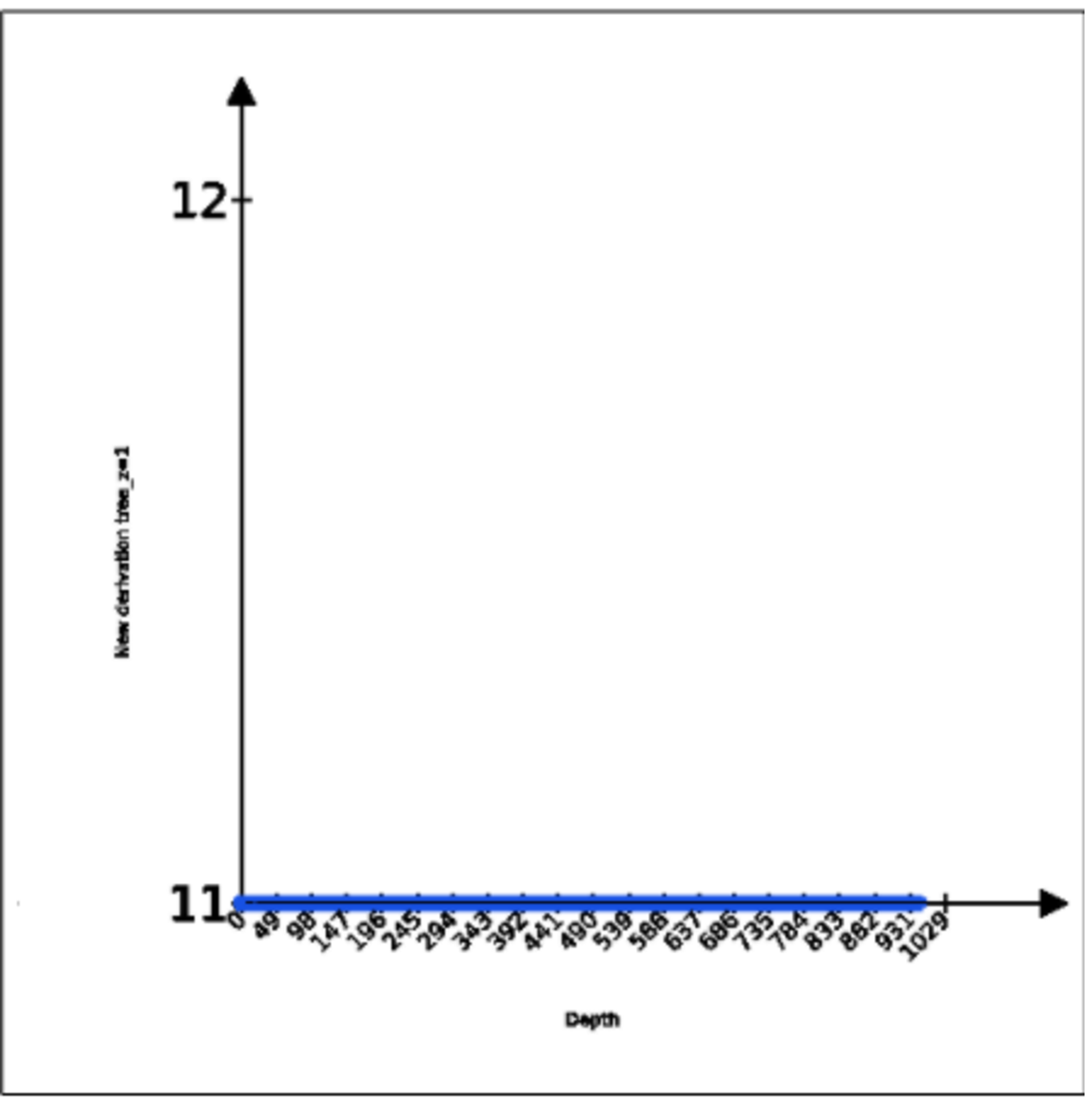}\label{figure:m1_ptox01_ddcost01}}
\subfloat[Low Toxicity, Low Due Diligence Cost, Negligent Banks]{\includegraphics[width=.25\textwidth]{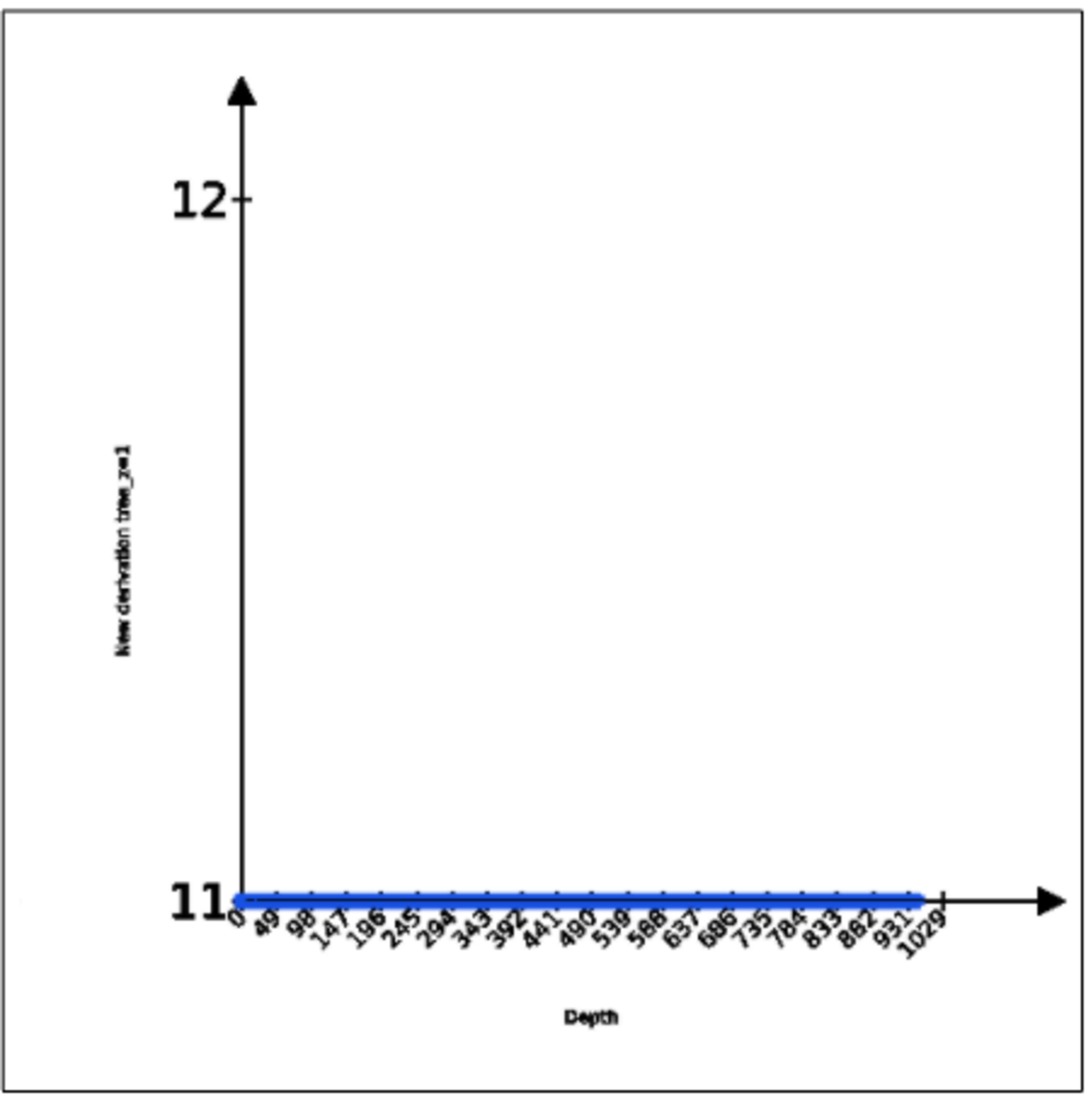}\label{figure:m1_ptox0001_ddcost0001}}
\subfloat[High Toxicity, Low Due Diligence Cost, Negligent Banks]{\includegraphics[width=.25\textwidth]{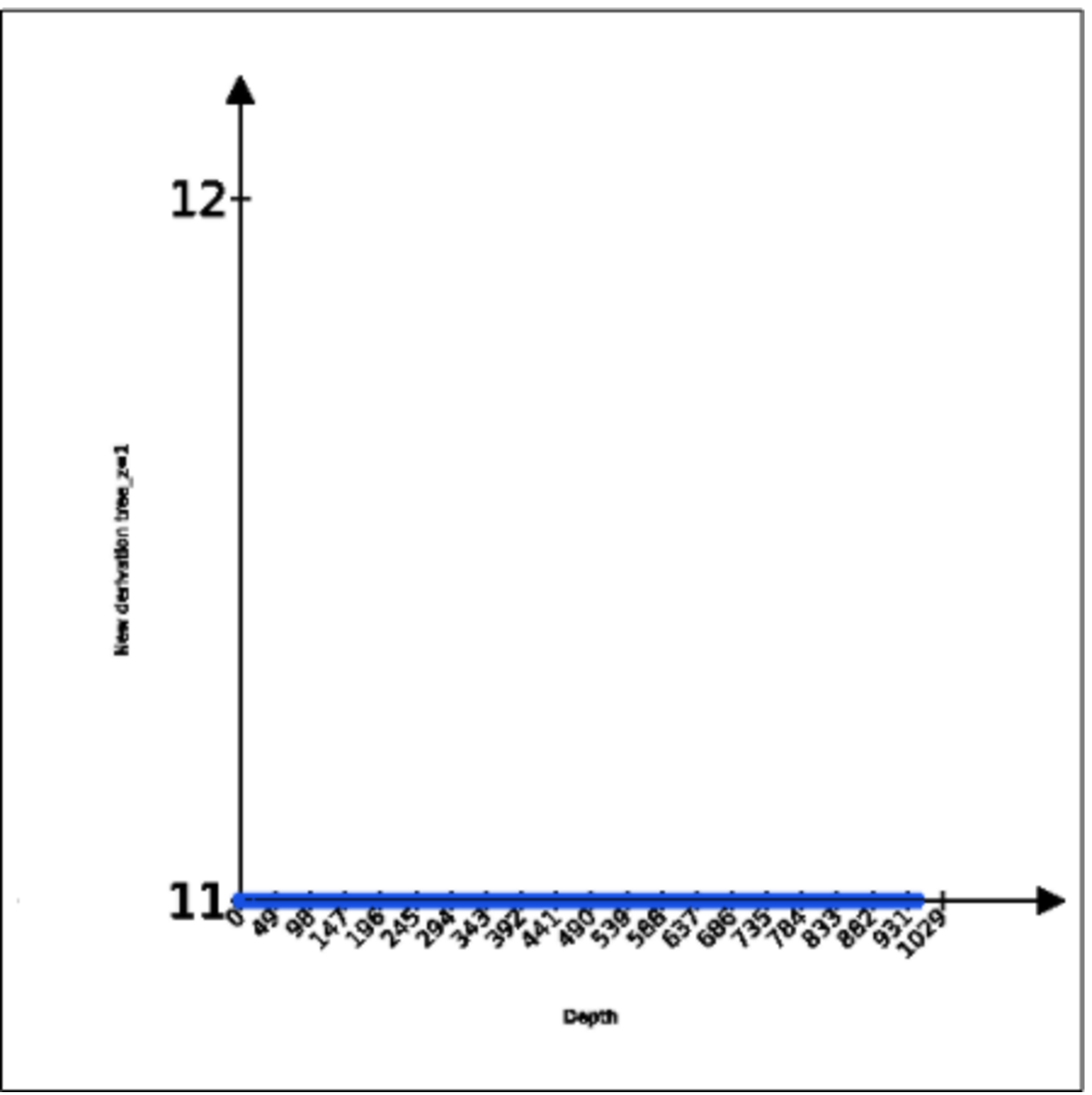}\label{figure:m1_ptox01_ddcost0001}}
\caption{\small \sl Experiment Results. \emph{(y-axis: Count of the number of negligent banks. The intersection of x and y axes in the case of a starting universe of purely diligent banks corresponds to the co-ordinates (0,0) as opposed to (11,0) in the case where we begin with negligent banks. Curves tending upwards reflect a negligent equilibrium result)}}
\label{figure:results}
\end{figure}

A successful base case validation has seen test results (see Figure~\ref{figure:results} where average Z count value is plotted versus depth of the simulation) line up qualitatively with results from the more traditional ABM simulation given in~\cite{FIN_agents_absmodel_}. In particular, for high values of $p$ (that is, high probability of toxicity), we observe the expected result when the initial state contains a mixture of negligent and diligent agents: a sharp drop in $Z$, corresponding to a sharp switch in average approach (i.e., more banks decide to perform independent analysis), which in turn will generate stability. An illustration of this can be seen in Figure~\ref{figure:m0545_ptox01_ddcost0001} and notice that given high due diligence costs Figures~\ref{figure:m0545_ptox01_ddcost01} and~\ref{figure:m0_ptox01_ddcost01} highlight a negligent approach whereas Figures~\ref{figure:m0545_ptox01_ddcost0001} and~\ref{figure:m0_ptox01_ddcost0001} reflect the favouring of a diligent approach. However, even for high toxicity, if the initial state is a set of negligent agents, the model reaches equilibrium without switching approach as seen in Figure~\ref{figure:m1_ptox01_ddcost0001}.

\section{Some Related Work and Conclusion}
\label{section_conclusion}
Fundamentally, general purpose agent-based simulation tools and platforms\footnote{\url{http://jasss.soc.surrey.ac.uk/18/1/11.html}} like JAS, Netlogo, AgentBuilder, Swarm, MASON, Repast, SeSAm, GAMA and INGENIAS Development Kit, support an imperative object-oriented approach to model development, facilitating the modular approach to coding. EMERALD and JADE middleware integrate a declarative approach but without any visualization support. Other tools and languages like Stratego, Maude and ELAN~\cite{MARTIOLIET2005417} support a pure term-writing approach which in the case of Maude is augmented by probabilistic features. What can we do in our chosen approach that we could not have achieved using existing solutions cited? The visual, declarative nature of graph transformation systems are welcome in the cases where users seek to primarily focus on describing what the system should accomplish in terms of problem domain versus the how, and maintain strong conceptualization support that can subsume the details of spatial and topological constraints.

We have shown that strategic port-graph rewriting provides a basis for the design and implementation of a multi-level graph model able to capture the inner workings of the sub-prime secondary securitisation market in a manner that reflects the aforementioned rational negligence phenomenon and that provides optional operational support. 

We observed that a declarative approach is much easier to program and maintain, and the incremental manner in which development was approached (e.g. coarse-grained rules tested before finer optimizations), in addition to the modular nature of development, eliminated many coding bugs. Interacting with the system was not convoluted in anyway and being able to view the states generated by each rule within the resulting derivation tree was useful in addition to being able to view nodes within the tree that maintained specific properties.   

~In future, we hope to further develop the hierarchical model to be able to capture all details of the full securitisation life-cycle and cater for more dynamically to changing parameters. 

\bibliographystyle{eptcs}
\bibliography{bibentries}

\end{document}